\documentclass[twocolumn,prb,amsmath,amssymb,floatfix]{revtex4-1}
\usepackage{amsmath,amssymb,color}
\usepackage{bm}
\usepackage{bbm}
\usepackage{graphicx}
\usepackage{psfrag}
\usepackage{ulem}
\usepackage{hyperref}
\usepackage{multibib}
\newcommand{\bd}{\bm}
\makeatletter
\@addtoreset{equation}{myequation}
\makeatother

\hypersetup{colorlinks=true,linkcolor=[rgb]{.85,.25,.31},citecolor=[rgb]{.17,.33,.66},urlcolor=[rgb]{.17,.33,.66}}

\begin{document}

\title{Zero-magnon sound in quantum Heisenberg 
ferromagnets}

\author{Raphal Goll, Andreas R\"{u}ckriegel, and Peter Kopietz}
  
\affiliation{Institut f\"{u}r Theoretische Physik, Universit\"{a}t
  Frankfurt,  Max-von-Laue Stra{\ss}e 1, 60438 Frankfurt, Germany}

\date{October 22, 2020}

 \begin{abstract}
Using a functional renormalization-group approach, we show that at low temperatures and in a certain range of magnetic fields, the longitudinal dynamic structure factor of quantum Heisenberg ferromagnets in dimensions $D \leq 2$ exhibits a well-defined quasiparticle peak with linear dispersion that we identify as zero-magnon sound.
In $D > 2$, the larger phase space available for the decay into transverse spin waves leads only to a broad hump centered at zero frequency whose width scales linearly in momentum.
\end{abstract}


\maketitle


\section{Introduction}
\label{sec:introduction}
At low temperatures, the transverse spin dynamics of ordered Heisenberg magnets can be explained in terms of transverse spin waves (magnons) forming a  weakly interacting gas of quasiparticles~\cite{Keffer66}.
Such a simple physical picture is not available for the longitudinal spin dynamics, which is a subject of ongoing research~\cite{Krivoruchko17,Krivoruchko20,Rodriguez18}.   
Perturbative spin-wave theory based on an  expansion in powers of the inverse spin quantum number $1/S$ fails in this case because the longitudinal spin dynamics, 
encoded in the longitudinal dynamic structure factor $S^{zz} ( \bd{q}, \omega )$, 
is dominated by emergent collective modes such as diffusion or sound modes.
Depending on the timescale of interest, 
two regimes should be distinguished: if the relevant timescale is large compared with the typical time between collisions, we enter the collision-dominated {\it{hydrodynamic regime}}
where the general form of 
 $S^{zz} ( \bd{q}, \omega )$ for small momenta $\bd{q}$ and frequencies $\omega$ can be obtained from hydrodynamic equations which follow from the continuity equations of the conversed quantities~\cite{Reiter68,Halperin69,Schwabl70}. 
On the other hand, in the {\it{collisionless regime}}, the timescale of interest is short compared to the typical collision time. 
The hydrodynamic approach is then not valid and microscopic calculations are  necessary.
\\
\indent
In spite of the long history of the problem, 
there is still no general agreement on some aspects of the longitudinal spin dynamics in Heisenberg ferromagnets, especially in reduced dimensions.
In the hydrodynamic regime,
the dynamics of the conserved quantities was obtained in the late 1960s by several authors \cite{Reiter68,Halperin69,Schwabl70}.
These studies found that at low temperatures and in the presence of a finite external magnetic field,
$S^{zz}(\bd{q},\omega)$ exhibits a diffusive peak at $\omega=0$ and a damped propagating sound mode, called \textit{second magnon}, in analogy with the second-sound mode in phonon systems.
When the temperature is increased, Umklapp scattering overdamps this mode and diffusive and propagating modes merge into a single broad diffusive peak.
A few years later, the problem was reconsidered by Dewel \cite{Dewel76,Dewel77}, 
who concluded that the hydrodynamic description is only valid for external magnetic fields $H$ exceeding a threshold of the order of $\bd{q}^2$; for $H\rightarrow 0$, the gapless nature of the transverse magnons induces singularities in the collision integrals of the kinetic equations, which invalidate the assumptions of hydrodynamics and the system enters the collisionless regime.
More  recently, Rodriguez-Nieva {\it{et al.}}~\cite{Rodriguez18} analyzed the 
hydrodynamic equations
in two dimensions 
and concluded that for sufficiently strong magnetic fields
the longitudinal structure factor
exhibits a weakly damped second-magnon mode.
\\
\indent
The literature on the  collisionless regime is less consistent.
Early studies \cite{Izuyama69,Natoli70,Huber71,Harris71} focused on the
question of whether 
the longitudinal structure factor of a three-dimensional Heisenberg ferromagnet
exhibits a  {\it{zero-magnon}} mode 
which could possibly be generated by 
coherent creation and annihilation processes of transverse magnons.
This zero magnon can also be viewed as a collective fluctuation of the
magnon density and  is analogous  to the zero-sound mode  
of interacting fermions \cite{Pines89,Aldrich76}. 
Original claims \cite{Izuyama69,Natoli70} of the existence of a zero-magnon mode
for momenta at the boundary of the Brillouin zone were rejected as an artifact of an inaccurate mapping of the Heisenberg model onto an effective  
bosonic Holstein-Primakoff system \cite{Harris71}. 
Subsequently, a number of works relying on various 
fermionic or bosonic representations of the Heisenberg Hamiltonian
calculated the longitudinal spin susceptibility 
in two \cite{Huber71} and three dimensions \cite{Reinecke74,Natoli74,Dewel76,Rudoy05} 
within the 
random phase approximation (RPA). 
None of these works found any evidence of a well-defined finite frequency collective mode 
anywhere in the Brillouin zone. 
The problem was picked up again by 
Izyumov {\it{et al.}}~\cite{Izyumov02}, 
who used a sophisticated diagrammatic approach for quantum spin systems developed by
Vaks \textit{et al.} \cite{Vaks68a,Vaks68b} to calculate the
longitudinal structure factor of a three-dimensional Heisenberg ferromagnet within a generalized RPA.
Unfortunately, the results for the longitudinal structure factor 
are somewhat inconclusive:
While their spectral line shape reveals broad finite-frequency maxima which scale linearly with $| \bd{q} |$, it is equally plagued by strange features
whose origin can be traced back to the gapless nature of the magnons in the limit $H\rightarrow 0$ (cf.~Ref.~[\onlinecite{Dewel77}]).
\\
\indent
In this work, we use the recent advances in the application of the functional renormalization group (FRG) to (quantum) spin systems \cite{Machado10,Rancon14,Krieg19,Goll19} to settle the longstanding question of the existence of a zero-magnon mode in quantum Heisenberg ferromagnets. 
Using an RPA truncation of the flow equations,
we show that a well-defined zero-magnon mode can exist in a certain range of magnetic fields in dimensions $D\le 2$ in the collisionless regime.

\section{Spin FRG}
The spin FRG approach \cite{Krieg19,Goll19} combines the advantages of working with  
physical spin operators with the well-known diagrammatic 
structure of the FRG vertex 
expansion \cite{Kopietz10,Metzner12,Dupuis20}, thus avoiding
the diagrammatic complexity inherent in  the spin-diagram 
technique \cite{Vaks68a,Vaks68b,Izyumov02,Izyumov88}.
Let us outline the main features of the method 
for the specific case of an anisotropic quantum Heisenberg ferromagnet with Hamiltonian
 \begin{equation}
 {\cal{H}} = - H \sum_{i} S^z_i - \frac{1}{2} \sum_{ij} 
 \left[ J^\bot_{ij} \bd{S}^{\bot}_i \cdot \bd{S}^{\bot}_j 
 + J^z_{ij} S^z_i S^z_j \right] ,
 \label{eq:hamiltonian}
 \end{equation}
where $i, j = 1, \ldots , N$ label the sites of a $D$-dimensional hypercubic lattice,
$\bd{S}_i = ( S_i^x, S_i^y , S^z_i ) = ( \bd{S}_i^{\bot} , S^z_i )$ are spin-$S$ operators,
the external magnetic field $H$ is measured in units of energy, 
and $J^\alpha_{ij} > 0$ (where $\alpha = \bot, z$) are 
ferromagnetic exchange couplings.
Following  Refs.~[\onlinecite{Krieg19}] and [\onlinecite{Goll19}],
we replace the exchange couplings $J^{\alpha}_{ij}$
in Eq.~(\ref{eq:hamiltonian}) by deformed couplings
$J^{\alpha}_{ \Lambda , ij}$ that depend
on a continuous parameter
$\Lambda \in [ 0, 1]$ such that 
$J^{\alpha}_{\Lambda =1, ij} = J^{\alpha}_{ij}$ and $J^{\alpha}_{\Lambda =0, ij}$ is 
simple enough to allow for a controlled solution of the model. 
In this work, we choose  
$J^{\bot}_{\Lambda  , \bd{k}} = \Lambda J^{\bot}_{\bd{k}}$ and 
$J^z_{\Lambda, \bd{k}} = J^{z}_{\bd{k}}$, 
so that at the initial value 
$\Lambda =0$, the transverse exchange interaction is completely switched off while the longitudinal interaction is not modified.
\\
\indent
Given such a continuous deformation, 
it is possible to derive a formally exact hierarchy of FRG flow equations for the imaginary-time ordered spin correlation functions \cite{Krieg19,Goll19}. 
\begin{figure}[tb]
 \begin{center}
  \centering
 \includegraphics[width=0.475\textwidth]{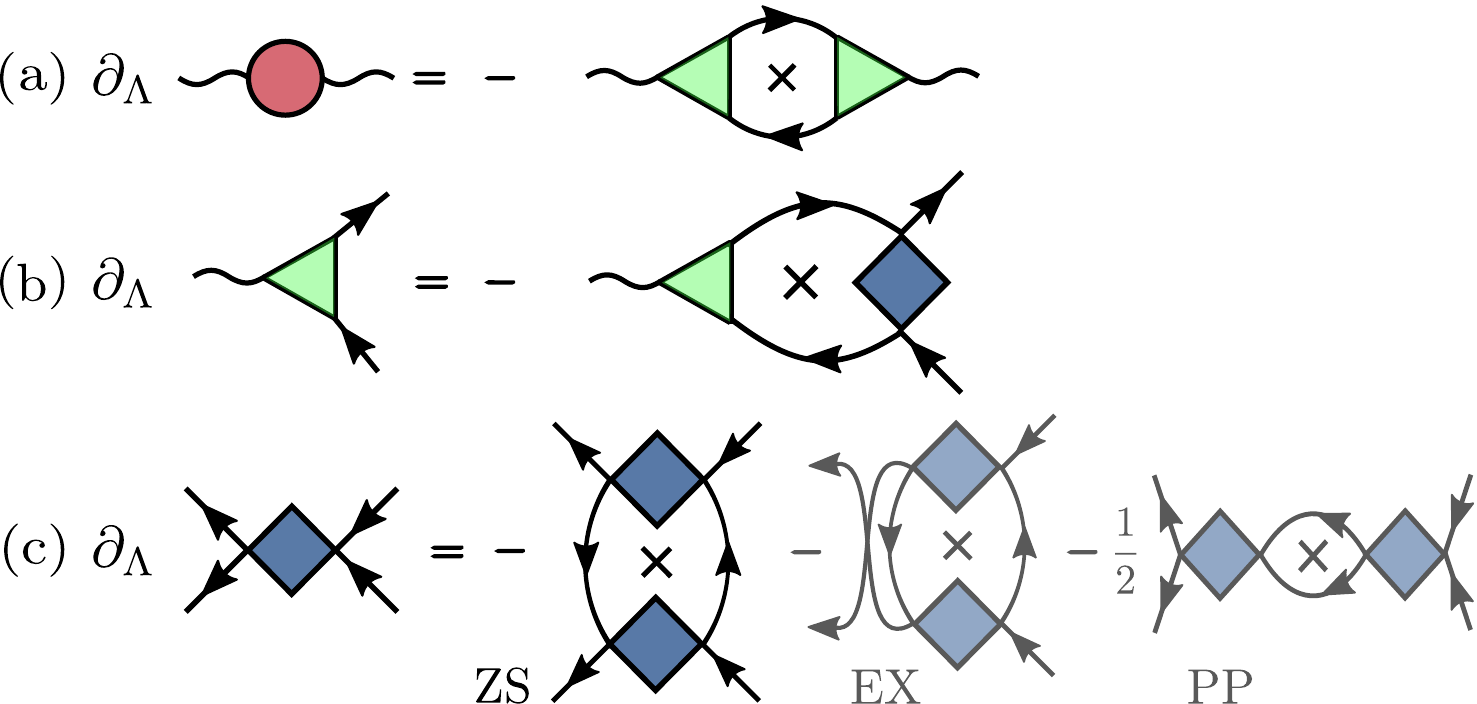}
   \end{center}
  \vspace{-5mm}
  \caption{
Diagrammatic representation of FRG  flow equations for (a)
the longitudinal two-point vertex, (b) the mixed three-point vertex, and (c) the
transverse four-point vertex appearing in the  truncated functional
$\tilde{\Gamma}_{\Lambda} [ \bd{m} , \varphi ]$ in Eq.~(\ref{eq:vertexp}).
In (c), we retain only the zero-sound (ZS) channel and neglect the exchange (EX) and particle-particle (PP) contributions.
Solid arrows represent transverse propagators, wavy legs represent the longitudinal exchange field $\varphi$, and the cross inside the loops means that each of the loop propagators is successively replaced by the corresponding single-scale propagator \cite{Kopietz10}.
}
\label{fig:diagrams}
\end{figure}
In the magnetically ordered phase, 
it is, however, more convenient to consider the flow of the functional $\tilde{\Gamma}_{\Lambda} [ \bd{m} , \varphi] $,
which depends on the transverse magnetization $\bd{m}$ and the fluctuating part $\varphi$ of the longitudinal exchange field
and generates vertices which are irreducible 
with respect to cutting a single transverse propagator line and a single longitudinal interaction line.
The explicit construction of this functional via a subtracted Legendre transformation of the generating functional of connected spin corrrelation functions, and the general structure of the vertex expansion of $\tilde{\Gamma}_{\Lambda} [ \bd{m} ,  \varphi ]$ have been discussed in Ref.~[\onlinecite{Goll19}]. 
In accordance with the studies presented in Sec.~\ref{sec:introduction}, we assume in the following that the longitudinal spin dynamics are dominated by multiple creation and annihilation processes of transverse spin waves.
Within this assumption, it is then sufficient to use the following truncation:
 \begin{eqnarray}
   \tilde{\Gamma}_{\Lambda} [ \bd{m} ,  \varphi ]   & \approx &  
\int_K   \left[ \Gamma_{\Lambda}^{+-} ( K ) m^-_{-K} m^+_K
  + 
  \frac{1}{2!} \Gamma^{zz}_{\Lambda} ( K )  \varphi_{-K} \varphi_K  \right]
 \nonumber
 \\
 &  &  \hspace{-9mm} + \frac{1}{3!} \int_{K_1} \int_{K_2} \int_{K_3} 
 \delta ( K_1 + K_2 + K_3 ) 
 \nonumber
 \\
 & & \hspace{-7mm} \times 
 \Gamma^{+-z}_{\Lambda}(K_1, K_2, K_3 ) m^-_{K_1} m^+_{K_2} \varphi_{K_3} 
 \nonumber
 \\
 &  &  \hspace{-9mm} + \frac{1}{(2!)^2} \int_{K_1} \int_{K_2} \int_{K_3} \int_{K_4} 
 \delta ( K_1 + K_2 + K_3 + K_4 ) 
 \nonumber
 \\
 & & \hspace{-7mm} \times 
 \Gamma^{++--}_{\Lambda}(K_1, K_2, K_3, K_4) m^-_{K_1} m^-_{K_2} m^+_{K_3} m^+_{K_4}. 
 \label{eq:vertexp}
 \end{eqnarray}
Here, 
$K = ( \bd{k} , i \omega )$ represents momentum and bosonic Matsubara frequency,
$\int_K = ( \beta N )^{-1} \sum_{\bd{k} , \omega }$ where $N$ is the number of lattice sites and
$\beta =1/T$ is the inverse temperature,  
$\delta ( K ) = \beta N \delta_{\bd{k} , 0} \delta_{\omega, 0}$, and
$m^{\pm}_K = ( m^x_K \pm i m^y_K ) / \sqrt{2}$ are the spherical Fourier components of the transverse magnetization.
\begin{figure*}[tb]
 \begin{center}
  \centering
 \includegraphics[width=1.0\textwidth]{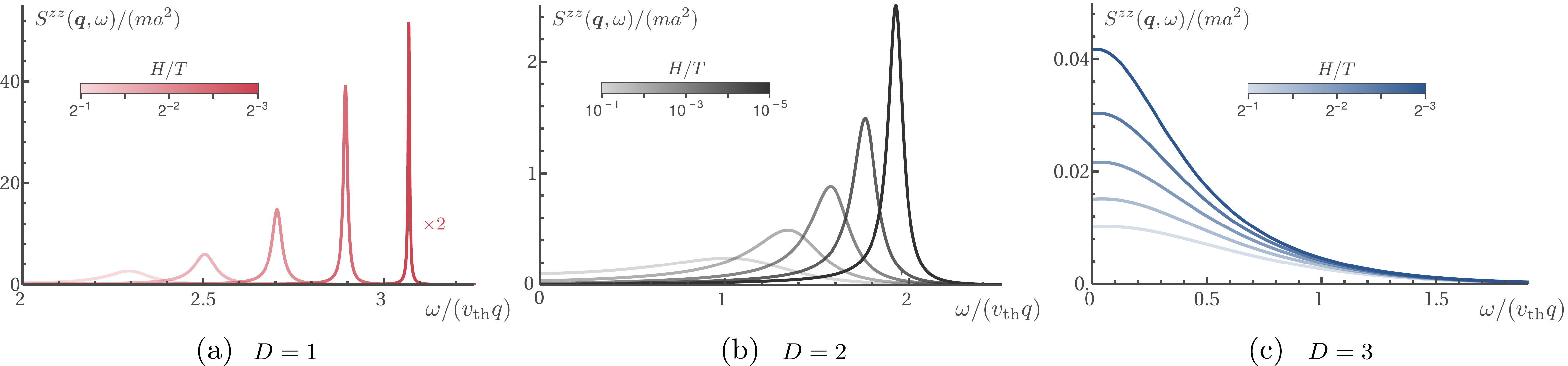}
   \end{center}
\vspace{-5mm}
  \caption{
  \small{
Longitudinal dynamic structure factor $S^{zz}(\bd{q},\omega)$ [Eq.~(\ref{eq:dynre})] 
of the $D$-dimensional Heisenberg ferromagnet in the regime $H <  T \ll 1/(ma^2) $, 
for $q=k_{\text{th}}/4$, temperature $T m a^2 =10^{-2}$, and $S=1/2$.
}
}
\label{fig:dynstruc}
\end{figure*}
The transverse two-point vertex $\Gamma_{\Lambda}^{+-} ( K )$ is related to the flowing propagator of transverse magnons \cite{Goll19},
\begin{equation}
G_{\Lambda} ( K )  = 
1/\left[ \Gamma^{+-}_{\Lambda} ( K ) + J^{\bot}_{\bd{k}} - J^{\bot}_{\Lambda , \bd{k}} \right],
\end{equation}
where $J^{\alpha}_{\bd{k}}$ is the Fourier transform of $J^{\alpha}_{ij}$. 
The longitudinal two-point vertex,
\begin{equation}
\Gamma^{zz}_{\Lambda} ( K ) = 1/ J^z_{\bd{k}}  - \Pi_{\Lambda} ( K ),
\end{equation}is related to the 
interaction-irreducible polarization $\Pi_{\Lambda} ( K )$ \cite{Goll19},
which in turn determines the flowing longitudinal spin susceptibility \cite{Vaks68a,Goll19}, 
\begin{equation}
\chi^{zz}_{\Lambda} ( K )  =   \Pi_{\Lambda} ( K ) / \left[ 1 - J^{z}_{\Lambda , \bd{k}}  \Pi_{\Lambda} ( K ) \right].
\label{eq:Gammazz}
\end{equation}

\section{Zero-sound truncation}
We are only interested in the longitudinal two-point vertex 
in the magnetically ordered regime for sufficiently low temperatures and high frequencies, 
i.e., in the collisionless regime. 
Therefore, we can neglect the flow of the magnetization and of the transverse two-point vertex.
In this approximation, we obtain
\begin{equation}
\Gamma^{+-}_{\Lambda} ( K ) \approx 
\Gamma_0^{+-} ( K ) = G_0^{-1} ( i \omega ) - J^{\bot}_{\bd{k}},
\end{equation}
where 
\begin{equation}
G_0 ( i \omega )=S/[ H +  J^z_0 S-i\omega].
\end{equation}
Within our truncation the three remaining vertices in Eq.~(\ref{eq:vertexp})
then satisfy the flow equations shown diagrammatically in Fig.~\ref{fig:diagrams}.
Explicit analytical expressions for the corresponding flow equations
can be found in Appendix A.
Note that the flow of the transverse four-point vertex in Fig.~\ref{fig:diagrams}~(c)
is driven by three different scattering channels, 
which we label ZS (zero sound), EX (exchange), and PP (particle-particle)  \cite{Kopietz10,Shankar94}.
To understand the collective modes of the longitudinal structure factor,
we focus in this work on the ZS channel and neglect the EX and PP channels.
For interacting fermions, this approximation amounts to the RPA and can be 
formally justified if the interaction is dominated by small momentum transfers. 
We therefore expect that also in the present case, the ZS channel dominates the renormalization of the effective interaction if $J^z_{\bd{k}}$ is enhanced for small $\bd{k}$. 
We assume that this ZS truncation remains at least qualitatively correct in the collisionless regime, even for short-range interactions.
\begin{figure*}[tb]
 \begin{center}
  \centering
 \includegraphics[width=1.0\textwidth]{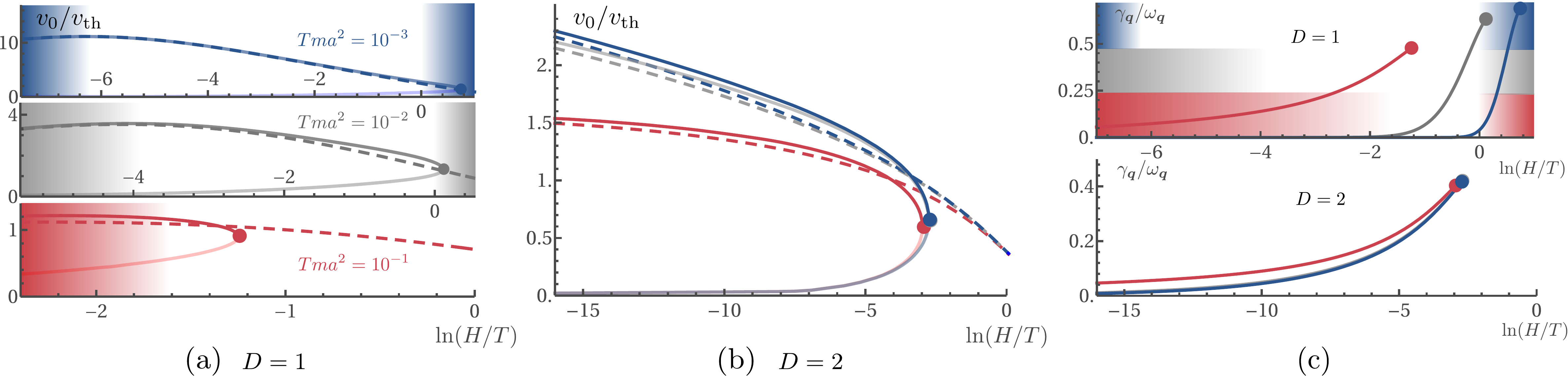}
   \end{center}
  \vspace{-5mm}
  \caption{
  \small{
Zero-magnon velocity $x_0=v_0 / v_{\rm th}$
in (a) $D=1$ and (b) $D=2$ as a function of dimensionless magnetic field $H/T$ 
for temperatures $T m a^2=10^{-1},10^{-2},10^{-3}$ (red, gray, blue).
The dashed line corresponds to the approximate solution \eqref{eq:xres}, 
while the full line is obtained solving Eq.~\eqref{eq:xroot} numerically. 
For magnetic fields smaller than a certain upper limit (displayed as a full dot), Eq.~\eqref{eq:xroot} has two solutions [lower (opaque) and upper (full) branch], 
where the larger one may be identified as the corresponding zero-magnon velocity.
The shading indicates the boundaries $H=H_*(T)$ and $H=T$ where our calculation breaks down. 
In $D=2$, the calculation remains valid for the whole range of displayed magnetic fields.
(c): Damping $y_0 = \gamma_{\bd{q}} / \omega_{\bd{q}}$ of the zero-magnon mode. 
}
}
\label{fig:velocity}
\end{figure*}
In the region $T \ll J_0^z S$, the resulting system of truncated flow equations can be solved 
analytically, which is shown explicitly in Appendix B.
The final result for the longitudinal spin susceptibility is \cite{note1}
 \begin{eqnarray}
 & & \chi^{zz} ( Q )   =
 \nonumber
 \\
 & &  \frac{ P^{00} ( Q ) }{ [ 1 + P^{01} ( Q ) ][1 + P^{10} ( Q ) ] - P^{00} ( Q ) [ P^{11} ( Q ) +  J^z_{\bd{q}} ] },
 \hspace{7mm}
 \label{eq:chire}
 \end{eqnarray}
with the generalized  polarization functions
 \begin{subequations}
 \label{eq:Polres}
 \begin{eqnarray}
S^2 {P}^{00} ( Q ) & = & \int_K L ( K,Q) ,
 \\
 S^2 {P}^{01} ( Q ) & = & S^2 {P}^{01} ( - Q ) = \int_K L ( K,Q) G_0^{-1} ( i \omega ),
 \\
 S^2 {P}^{11} ( Q ) & = & \int_K L ( K,Q) G_0^{-1} ( i \omega ) G_0^{-1} ( i \omega - i \nu).
 \end{eqnarray}
 \end{subequations}
Here, 
$i \nu $ is the Matsubara frequency of the external label
$Q = ( \bd{q} , i \nu )$, and the function $L ( K,Q) $ is defined by
 \begin{equation}
 L ( K, Q ) = {G} ( K ) G ( K - Q )   - G_0 ( i\omega  ) G_0 ( i \omega - i \nu ).
 \end{equation}
Here,
\begin{equation}
G ( K ) = S/[ H + \epsilon_{\bd{k}} - i \omega ],
\end{equation}
is the propagator of transverse magnons with dispersion
$\epsilon_{\bd{k}} =  S ( J_0^z - J_{\bd{k}}^{\bot} )$.  
\\
\indent
To make progress analytically, 
we now assume an isotropic exchange interaction $J^{\bot}_{\bd{k}} = J^{z}_{\bd{k}} = J_{\bd{k}}$
and focus on the regime $H \lesssim  T \ll 1/(2 m a^2)$, where $a$ is the lattice spacing 
and the magnon mass $m$ is defined via the small-momentum expansion of the 
magnon dispersion, $\epsilon_{\bd{k}} = \bd{k}^2/(2m)$.
Note that at low temperatures
the thermal momentum
$k_{\rm th} = \sqrt{2 m  T } \ll 1/a $ acts as ultraviolet cutoff  for all momentum integrals,
which justifies the small-momentum expansion of the magnon dispersion.
For $q \ll k_{\rm th}$ we may then expand the polarization functions
in Eq.~(\ref{eq:chire}) in powers of $q$. 
Neglecting terms of the order of $q^3$ and higher, we obtain 
\begin{equation}
P^{00} ( \bd{q} , i \omega ) = P ( i \omega / ( v_{\rm th} q) ),
\end{equation}
where $v_{\rm th} = k_{\rm th}/m$
is the thermal velocity, 
and for complex $z$, the function $P(z)$ is in $D$ dimensions given by
  \begin{eqnarray}
  P ( z  )  & = &  \frac{\Omega_D ( k_{\rm th}  a )^D}{( 2 \pi )^D 2 T}
 \int_0^{\infty} d\epsilon 
 g_D  \left( \frac{ z  }{ \sqrt{\epsilon} } \right)
\frac{  \epsilon^{\frac{D-2}{2}}     e^{h + \epsilon }}{ [e^{ h + \epsilon } -1]^2 }.
 \hspace{9mm}
 \label{eq:Pin}
 \end{eqnarray}
Here, $\Omega_D = 2 \pi^{D/2}/ \Gamma ( D/2)$ is the surface of the $D$-dimensional unit sphere [with the Gamma-function $\Gamma(z)$],
$h = H /T$ is the dimensionless magnetic field,
and the function $g_D ( z )$ is defined via the $D$-dimensional angular 
average \cite{Kopietz97},
\begin{equation}
g_D ( z ) = \frac{1}{\Omega_D} \int d \Omega  \frac{ \cos \vartheta }{ \cos \vartheta  - z },
\label{eq:angularint}
\end{equation}
where the angle $\vartheta$ is the latitude on the surface of the unit sphere.
Moreover, to leading order in $\bd{q}$, the other polarization functions
in Eq.~(\ref{eq:chire})  are given by
\begin{subequations}
\begin{align}
P^{01} ( \bd{q} , i \omega ) &= P^{10} ( \bd{q} , i \omega ) = J_0 P^{00} ( \bd{q} , i \omega ) + \rho,
\\
P^{11} ( \bd{q} , i \omega ) &= J_0^2 P^{00} ( \bd{q} , i \omega ) + 2 J_0 \rho,
\end{align}
\end{subequations}
%
where
 \begin{equation}
 \rho = \frac{1}{SN} \sum_{\bd{k}} \frac{1}{e^{(  H + \epsilon_{\bd{k}} )/T} -1 }
 \label{eq:magdens}
 \end{equation}
is the density of thermally excited magnons in units of the saturated magnetization $S$.
As described in detail in Appendix B,
our result (\ref{eq:chire}) for  the longitudinal susceptibility then reduces to
\begin{equation}
 \chi^{zz} ( \bd{q} , i \omega ) = \frac{P (  \frac{i \omega }{ v_{\rm th} q } ) }{
 ( 1 + \rho )^2 + J_0  P (  \frac{i \omega}{ v_{\rm th} q } ) }.
 \label{eq:chire2}
 \end{equation}

\section{Zero-magnon sound}
To investigate the existence of a collective zero-magnon mode,
we now consider the longitudinal dynamic structure factor,
\begin{equation}
S^{zz} ( \bd{q} , \omega )  =   
\left[ 1 + \frac{1}{ e^{\omega/T} -1 } \right] 
\frac{1}{\pi} {\rm Im} \chi^{zz} ( \bd{q} , \omega + i 0^+ ) ,
\label{eq:dynre}
\end{equation}
which can be obtained by evaluating the integral defining the 
function $P ( z)$ in Eq.~(\ref{eq:Pin}) numerically. 
The explicit expressions for $P ( z)$ in $D=1,2,3$ can be found in Appendix C and the resulting line shapes 
are shown in Fig.~\ref{fig:dynstruc}. 
In one and two dimensions, we observe a finite-frequency peak at position $\omega_{\bd{q}} = v_0 q$.
In $D=1$, the peak is sharp as soon as the magnetic field is only slightly smaller than the temperature, such that for $H \lesssim T$, we can identify $\omega_{\bd{q}}$ as the zero-magnon mode.
In the two-dimensional case, the peak broadens and a similar well-defined mode is only obtained for significantly smaller magnetic fields.  
In contrast, in $D=3$, the longitudinal structure factor exhibits only a broad hump centered at $\omega=0$, whose width is proportional to $q$. This indicates that there is no zero-magnon mode in this case.
\\
\indent
To understand the origin of these results,  
we note that  a well-defined zero-magnon peak in $S^{zz} ( \bd{q} , \omega ) $ exists 
if the susceptibility (\ref{eq:chire2}) has a pole close to the real axis in the complex frequency plane.
The dispersion of the
zero-magnon mode is then $\omega_{\bd{q}} = v_0 q = x_0 v_{\rm th} q$, where the real number 
$x_0$ 
is the positive root of the equation
 \begin{equation}
 ( 1 + \rho )^2 + J_0 {\rm Re} P ( x_0 + i0^+ ) =0.
 \label{eq:xroot}
 \end{equation}
Expanding around $x_0 $, we obtain a Lorentzian line shape
for the longitudinal structure factor in the vicinity of the zero-magnon mode,
\begin{equation}
S^{zz} ( \bd{q} , \omega )  = 
\left[  1 + \frac{1}{e^{\omega/T} -1 } \right] \frac{\omega_{\bd{q}}}{ 2 \pi J_0}    \frac{  \gamma_{\bd{q}}}{ ( \omega
- \omega_{\bd{q} } )^2 + \gamma_{\bd{q}}^2}.
\end{equation} 
The damping is $\gamma_{\bd{q}} = y_0 v_{\rm th} q $ where
$y_0 = {\rm Im} P ( x_0 + i 0^+ ) / {\rm Re} P^\prime ( x_0 + i 0^+ )$.
Numerical results for the root $x_0 = v_0 / v_{\rm th}$ of Eq.~(\ref{eq:xroot}) and the relative damping
$ \gamma_{\bd{q}} / \omega_{\bd{q}} = y_0 / x_0$ for $D=1,2$
are shown in Fig.~\ref{fig:velocity}.
We note, in particular, that the mode is only well defined ($\gamma_{\bd{q}} / \omega_{\bd{q}}\ll 1$) if the zero-sound velocity is significantly larger than the thermal velocity, i.e., if $x_0 \gg 1$.
For $D=3$, Eq.~(\ref{eq:xroot}) has no solutions, so that there is no well-defined zero-magnon mode in this case. 
\\
\indent
To gain an analytical understanding of these observations,
let us anticipate that for $D \leq 2$, the integral defining the function $P ( z )$ in Eq.~(\ref{eq:Pin}) is dominated by the regime $ \sqrt{\epsilon} \ll  | z |$, where we may approximate 
$g_D ( z / \sqrt{\epsilon} ) \approx - \epsilon / ( D z^2 )$ (see Appendix C).
This yields 
\begin{equation}
P ( z )  \approx  - \frac{  \left( k_{\rm th}  a \right)^D    }{ 2 c_D  T z^2}
       {\rm Li}_{ \frac{D}{2} } ( e^{ - h} ), \; \; \; \; \; c_D = 2^D \pi^{\frac{D}{2}},
\label{eq:Pintstrong}
\end{equation}
where ${\rm Li}_s ( z )$ is the polylogarithm.
The dimensionless magnon density (\ref{eq:magdens}) can likewise be written as
\begin{equation}
\rho = \frac{ ( k_{\rm th}  a )^D   }{ c_D S}
 {\rm Li}_{\frac{D}{2}} ( e^{ - h}).
\end{equation}
From Eq.~(\ref{eq:xroot}) we then obtain for the dimensionless velocity of the zero-magnon mode,
\begin{equation}
x_0 = \frac{ v_0}{v_{\rm th}} = \sqrt{ \frac{  \left( k_{\rm th}  a \right)^D J_0   
{\rm Li}_{ \frac{D}{2} } ( e^{ - h} ) }{ 2 c_D T ( 1 + \rho )^2 }}.
\label{eq:xres}
\end{equation}
Keeping in mind that the integral in Eq.~\eqref{eq:Pin} is cut for $\epsilon \lesssim 1$, we see that only for $x_0 \gg 1$, it is consistent to use the approximation \eqref{eq:Pintstrong}.
Since $( k_{\rm th}  a )^D\propto T^{D/2}$, this is not satisfied for $D>2$ at low temperatures,
so that there is no high-frequency zero mode.
In contrast, for $D \leq 2$ and at sufficiently small magnetic fields, there is always a parametrically large regime where $ x_0 \gg 1$, as shown in Fig.~\ref{fig:velocity}. 
Note, however, that our calculation is only applicable if the dimensionless magnon density $\rho$ is small compared with unity 
because we have assumed that the magnetization is almost saturated.
For $D \leq 2$, this implies that our results are only valid 
for magnetic fields $H \gg H_{\ast} \approx 1/(2 m \xi^2)$, 
where $\xi$ is the correlation length for $H=0$.
Using a one-loop approximation \cite{Takahashi87,Kopietz89},
we estimate 
$H_{\ast} \approx T e^{ - 2 \pi   S /(m a^2 T) }$ for  $D=2$, 
and $H_{\ast} \approx m a^2 T^2$ for $D=1$.
In the region $H_{\ast}\ll H \lesssim T$, the dimensionless magnon density $\rho$ in Eq.~\eqref{eq:xres} may thus be neglected and a well-defined zero-magnon mode is obtained in $D=2$ if $\ln (T/H) \gg 4 \pi /( ma^2 J_0)$.
In addition, it should be pointed out that our result is only applicable in the momentum regime $ 1/ \xi \ll q \lesssim  k_{\rm th}$
where the system appears to be magnetically ordered at the length-scale $1/q$ and
the expansion in powers of $q$ is valid.
\\
\indent
For a Heisenberg ferromagnet with nearest-neighbor exchange $J$, 
we estimate that the maximal value of the zero-magnon velocity (\ref{eq:xres})
is $v_0 \propto a J S$ for $H = H_{\ast}$ in $D=1,2$ (see Appendix C).
It is interesting to compare this 
with the velocity $v_2 \approx a \sqrt{ 2 J T {\rm Li}_2 ( z ) / {\rm Li}_1 ( z ) }$ 
of second-magnon sound in the hydrodynamic regime, 
where $z = e^{ -\beta \mu }$ is the magnon fugacity\cite{Rodriguez18}. 
Close to equilibrium, the magnon
chemical potential $\mu$ is small, so that   ${\rm Li}_2 ( z ) 
\approx {\rm Li}_2 ( 1 ) = \pi^2/6$ and ${\rm Li}_1 ( z ) \approx  \ln ( T / \mu )$, 
implying
$v_2 \ll v_{\rm th} \ll v_0$. 

\section{Conclusions}
In summary, we have uncovered the existence of a zero-magnon sound mode
in quantum Heisenberg ferromagnets in dimensions $D \le 2$, which
should be observable in one- and two-dimensional ferromagnets 
in the collisionless regime for $q \xi \gg 1$ and for magnetic fields in the range
$1/(2 m \xi^2) \ll H \lesssim T \ll 1/(2 ma^2)$.  
The signature of the zero magnon should be detectable with polarized neutron scattering \cite{Boeni91} and in the relaxation time of a spin qubit coupled to a ferromagnet, as discussed in Ref.~[\onlinecite{Rodriguez18}].
Finally, let us point out that our spin FRG approach can also be used to calculate the longitudinal spin dynamics of more general Heisenberg models, including antiferro- and ferrimagnets \cite{Krivoruchko20,Krivoruchko16}.

\section*{Acknowledgements}
We thank  J. F. Rodriguez-Nieva for useful correspondence.
This work was financially supported by the Deutsche Forschungsgemeinschaft (DFG)
through Project No. KO/1442/10-1.

\section*{APPENDIX A: Derivation of the truncated spin FRG flow equations}
\setcounter{equation}{0}
\renewcommand{\theequation}{A\arabic{equation}}

The truncated spin FRG flow equations shown graphically in Fig.~\ref{fig:diagrams} can be derived within the spin FRG scheme developed in Ref.~[\onlinecite{Goll19}].
The essential idea of this method is the asymmetric treatment of longitudinal and transverse fluctuations,
a procedure tailored to investigate spin systems in the ordered phase.
Technically, this is implemented via a hybrid generating functional $\tilde{\Gamma}_{\Lambda}[\bd{m},\varphi]$ 
that depends on the transverse magnetization $\bd{m}$ and on the fluctuations $\varphi$ around the longitudinal exchange field ${\phi}_{\Lambda}$.
The latter can be identified with the exchange correction to the external magnetic field. 
For the detailed construction of $\tilde{\Gamma}_{\Lambda}[\bd{m},\varphi]$ 
in terms of the generating functional  ${\cal{G}}_{\Lambda} [ \bd{h} ]$ of the time-ordered connected spin correlation functions, 
we refer to Ref.~[\onlinecite{Goll19}].
Similar to ${\cal{G}}_{\Lambda} [ \bd{h} ]$, 
the functional $\tilde{\Gamma}_{\Lambda}[\bd{m},\varphi]$ satisfies an exact flow equation which determines the evolution of the irreducible vertices 
as the interaction is gradually deformed. 
The explicit derivation (see Ref.~[\onlinecite{Goll19}]) yields
\begin{align}
 \partial_{\Lambda} \tilde{\Gamma}_{\Lambda} [ 
 \bd{m},  \varphi ] 
 &     =     
 \frac{1}{2} {\rm Tr} 
 \left\{ \left[
 \left( \mathbf{\tilde{\Gamma}}^{\prime \prime}_{\Lambda} [ \bd{m} , \varphi ] +
 \mathbf{R}_{\Lambda} \right)   - {\mathbf{J}}_{\Lambda}^z \right]  \partial_{\Lambda} \mathbf{R}_{\Lambda}    \right\}
 \nonumber
  \\
  & +   \left( \partial_{\Lambda} {\phi}_\Lambda \right)
   \int_K \delta(K) \frac{ \delta \tilde{\Gamma}_{\Lambda} [ 
 \bd{m} ,  \varphi  ]}{\delta {\varphi}_K },
 \label{seq:WetterichHeisenberg}
 \end{align} 
where the matrix elements of
$\mathbf{\tilde{\Gamma}}^{\prime \prime} _{\Lambda} [ \bd{m} , \varphi ]$
are  given by
 \begin{equation}
 \left(    \mathbf{\tilde{\Gamma}}^{\prime \prime} _{\Lambda} [ \bd{m} , \varphi  ]
 \right)_{K\alpha,K'\alpha'} =
 \frac{ \delta^2 \tilde{\Gamma}_{\Lambda} [ \bd{m} , \varphi  ] }{
 \delta \Phi^{\alpha}_K \delta \Phi^{\alpha^{\prime}}_{K'} }, 
  \end{equation}
with $\bd{\Phi}_K^T=(m_K^x,m_K^y,\varphi_K)$.
The regulator matrix $\mathbf{R}_{\Lambda}$ and the longitudinal exchange matrix $\mathbf{J}^z_{\Lambda}$ 
are diagonal in the field labels, 
with matrix elements
\begin{subequations}
 \begin{align}
[ \mathbf{R}_{\Lambda} ]^{xx}_{K,K' }
& =  [ \mathbf{R}_{\Lambda} ]^{yy}_{K,K' }
=   \delta(K-K')
R^\bot_\Lambda(\bd{k}),
 \\
[ \mathbf{R}_{\Lambda} ]^{zz}_{ K,K' }
 & =     \delta(K-K')
R^\phi_\Lambda(\bd{k}),
 \\
[ \mathbf{J}^z_{\Lambda} ]^{xx}_{ K,K' }
 & = [ \mathbf{J}^z_{\Lambda} ]^{yy}_{ K,K' }
 =  0,
 \\
[ \mathbf{J}^z_{\Lambda} ]^{zz}_{ K,K' }
 & =     \delta(K-K')
J^z_{\Lambda,\bd{k}} ,
 \end{align}
\end{subequations}
where the longitudinal and transverse regulators are given by
\begin{subequations}
\begin{align}
R^{\bot}_{\Lambda}(\bd{k}) =& J^{\bot}_{\bd{k}}-J^{\bot}_{\Lambda,\bd{k}},
\\
R^{\phi}_{\Lambda}(\bd{k}) =& \frac{1}{J^{z}_{\Lambda,\bd{k}}}-\frac{1}{J^{z}_{\bd{k}}}.
\end{align}
\end{subequations}
Here, $J_{\Lambda,\bd{k}}^{\bot}$ and $J_{\Lambda,\bd{k}}^{z}$ denote the Fourier transforms of the exchange couplings.
The specific deformation scheme is encoded in the $\Lambda$ dependence of the interaction.
In the present work, we chose a simple scheme, where the transverse interaction 
\begin{equation}
J^{\bot}_{\Lambda,\bd{k}}=\Lambda J^{\bot}_{\bd{k}}
\end{equation}
is continuously switched on with the help of a deformation parameter $\Lambda \in [0,1]$, while the longitudinal interaction is not deformed at all,
i.e.,
\begin{equation}
J^{z}_{\Lambda,\bd{k}}= J^{z}_{\bd{k}}.
\end{equation}
The hierarchy of flow equations for the vertices generated by $\tilde{\Gamma}_{\Lambda} [ 
 \bd{m},  \varphi ] $ can now be derived by substituting the ansatz defined in Eq.~\eqref{eq:vertexp} into the exact flow equation \eqref{seq:WetterichHeisenberg}.
The resulting equations are given by a sum of different loop integrals, 
where each loop can be classified according to the number of longitudinal propagators,
\begin{equation}
F_\Lambda(K)=\frac{J^z_{\bd{k}}}{1-J^z_{\bd{k}} \Pi_\Lambda(K)},
\end{equation}
which should be regarded as an effective screened interaction between the longitudinal spin fluctuations.
\\
\indent
In this work, we retain only the lowest-order contributions.
This corresponds to a zeroth-order expansion in the inverse interaction range \cite{Vaks68a,Vaks68b}, and,
similar to the RPA treatment of Fermi systems \cite{Pines89}, 
is thus formally justified only if $J^z_{\bd{k}}$ is enhanced for small $\bd{k}$. 
Within this approximation, 
we obtain the flow equations shown graphically in Fig.~\ref{fig:diagrams}.
The flow of the longitudinal polarization is given by
\begin{widetext}
\begin{subequations}
\label{seq:gammaflow}
\begin{equation}
\partial_\Lambda \Gamma^{zz}_\Lambda(Q)=
-\partial_\Lambda \Pi_\Lambda(Q)= 
- \int_K [G_\Lambda(K)G_\Lambda(K-Q)]^\bullet 
 \Gamma_\Lambda^{+-z}(K,K-Q,Q) \Gamma_\Lambda^{+-z}(K-Q,K,-Q),
\label{seq:gammazzflow}
\end{equation}
while the flow of the mixed three-point vertex reads,
\begin{equation}
\partial_\Lambda \Gamma_\Lambda^{+-z}(K+Q,K,Q) =
 -\int_{K'} [G_\Lambda(K)G_\Lambda(K+Q)]^\bullet 
 \Gamma_\Lambda^{+-z}(K'+Q,K',Q) \Gamma_\Lambda^{++--}(K+Q,K';K'+Q,K),
\label{seq:gamma3flow}
\end{equation}
and finally the flow of the transverse four-point vertex satisfies
\begin{align}
\partial_\Lambda \Gamma_\Lambda^{++--}(K_1+Q,K_2-Q;K_2,K_1) = 
&
-\int [G_\Lambda(K)G_\Lambda(K-Q)]^\bullet 
 \Gamma_\Lambda^{++--}(K_1+Q,K-Q;K,K_1) 
\nonumber\\
& \phantom{\int [G_\Lambda(K)G_\Lambda(K-Q)]^\bullet}
\times
\Gamma_\Lambda^{++--}(K_2-Q,K;K-Q,K_2).
\label{seq:gamma4flow}
\end{align}
\end{subequations}
\end{widetext}
Here, we introduced the abbreviation
\begin{align}
[G_\Lambda(K)G_\Lambda(K-Q)]^\bullet=
&
\dot{G}_\Lambda(K) G_\Lambda(K-Q) 
\nonumber\\
&
+  
G_\Lambda(K) \dot{G}_\Lambda(K-Q),
\label{seq:GGproduct}
\end{align}
where the transverse propagator and the corresponding single-scale propagator are defined as
\begin{subequations}
\label{seq:Propdef}
\begin{align}
G_\Lambda(K)
&= \frac{ 1}{ \Gamma_\Lambda^{+-}(K)+R^\bot_\Lambda(\bd{k}) } ,
\label{seq:Gdef}
\\[.2cm]
\dot{G}_\Lambda(K)
&=\left( \partial_\Lambda J_{\Lambda,\bd{k}}^{\bot} \right)G^2_\Lambda(K).
\label{seq:Gdotdef}
\end{align}
\end{subequations}
Note that in deriving the flow equation \eqref{seq:gamma4flow} for the transverse four-point vertex, 
we have neglected two further contributions corresponding to the additional exchange- and particle-particle diagrams, labeled by EX and PP in Fig.~\ref{fig:diagrams}.
In order to close the system of flow equations \eqref{seq:gammaflow}, 
we should, in principle, consider the flow of the transverse two-point vertex $\Gamma_\Lambda^{+-}(K)$ as well.
However, as explained in the main text, 
for our deformation scheme,
the associated self-energy corrections may be neglected in the ordered regime for sufficiently low temperatures, 
so that 
\begin{equation}
\Gamma_\Lambda^{+-}(K)\approx\Gamma_0^{+-}(K),
\end{equation}
which will be specified in Eq.~\eqref{seq:Gamma0def} below. 
The corresponding single-scale propagator may then be written as a scale derivative $\dot{G}_\Lambda(K)= \partial_\Lambda G_\Lambda(K)$, 
which is equivalent to the so-called Katanin substitution \cite{Katanin04}.
Eventually, 
we are thus left with the closed set of Eqs.~\eqref{seq:gammaflow}--\eqref{seq:Propdef}, 
which constitute the spin FRG analog of the Bethe-Salpeter equations derived and analyzed by Izyumov \textit{et al.} \cite{Izyumov02}.

\subsection*{Initial conditions}
In our deformation scheme,  the transverse interaction is initially switched off, while the longitudinal interaction is not modified. 
The deformed model at $\Lambda=0$ is then an Ising model, and the correlation functions cannot be calculated exactly. 
However, we are not interested in the critical regime, but rather focus on the low-temperature regime $T \ll J_0^z S$, so that a perturbative calculation of the correlation functions is possible. 
In particular, we consider the effect of the longitudinal exchange interaction only on a mean-field level 
(see Ref.~[\onlinecite{Goll19}] for a formal justification of this approximation and the explicit calculation of the initial connected correlation functions and irreducible vertices).
The resulting initial vertices are then given by
\begin{subequations}
\label{seq:Gamma0}
\begin{align}
\Gamma_{0}^{+-}(K)=&G_0^{-1}(i\omega)-J^\bot_{\bd{k}},
\label{seq:Gamma0def}
\\[.2cm]
\Gamma_{0}^{zz}(K)=&(J^z_{\bd{k}})^{-1} -\Pi_0(\omega),
\\[.2cm]
M_0 \Gamma_{0}^{+-z}(K_1,K_2,K_3)=&
1 -G_0^{-1}(i\omega_3)\Pi_0(\omega_3),
\label{seq:Gamma30def}
\\[.2cm]
M_0^2 \Gamma_{0}^{++--}(K_1,K_2;K_3,K_4&)=
G_0^{-1}(i\omega_3)+G_0^{-1}(i\omega_4)
\nonumber
\\
& \hspace{-3.8cm}
-
\left[\Pi_0(\omega_1-\omega_3)+\Pi_0(\omega_1-\omega_4)\right]
G_0^{-1}(i\omega_3)
G_0^{-1}(i\omega_4),
\label{seq:Gamma40def}
\end{align}
\end{subequations}
where the initial values of the inverse transverse propagator and the longitudinal polarization are
\begin{subequations}
\begin{align}
G_0^{-1}(i\omega)&=\frac{H+M_0 J^z_0-i\omega}{M_0},
\end{align}
and
\begin{align}
\Pi_0(\omega)&=\delta_{\omega,0} \beta b'(\beta (H +M_0 J^z_0 )).
\label{seq:Pi0def}
\end{align}
\end{subequations}
Here, $J^z_0=J^z_{\bd{q}=0}$, 
and the initial magnetization $M_0$ is determined self-consistently from
\begin{equation}
M_0  = b \left(\beta (H +M_0 J^z_0 )\right),
\end{equation}
where
\begin{align} 
  &b(y) = \left( S + \frac{1}{2} \right) \coth  \left[ \left( S + \frac{1}{2} \right) y \right] - \frac{1}{2} \coth \left[ \frac{y}{2} \right],
 \label{seq:brillouin}
 \end{align}
is the spin-$S$ Brillouin function and $b'(y)=\partial_y b(y)$ is its first derivative.
Using the relations \eqref{seq:Gdef} and \eqref{seq:Gamma0def}, we obtain, for the transverse propagator,
\begin{align}
G_{\Lambda}(K)
=&
\frac{1}{G_{0}^{-1}(K)-J^\bot_{\Lambda,\bd{k}}}
=
\frac{M_0}{H+\epsilon_{\Lambda,\bd{k}}-i\omega},
\end{align}
with the $\Lambda$-dependent magnon dispersion 
\begin{align}
\epsilon_{\Lambda,\bd{k}}=M_0 (J^z_0-J^\bot_{\Lambda,\bd{k}}).
\end{align}

\section*{APPENDIX B: Generalized RPA solution of the longitudinal dynamic spin susceptibility}
\setcounter{equation}{0}
\renewcommand{\theequation}{B\arabic{equation}}
In this appendix, 
we derive the generalized RPA solution given in Eq.~\eqref{eq:chire}, 
starting from the set of FRG flow equations \eqref{seq:gammaflow}.
Given the structure of these equations, 
it is clear that in order to find an explicit solution for the polarization, 
we have to choose a suitable parametrization of the three- and four-point vertices. 
In the following, we focus on the low-temperature regime, $T\ll J_0^zS$.
The initial magnetization can then be approximated by $M_0\approx S$ 
and we can furthermore neglect the exponentially small terms involving derivatives of the Brillouin function. 
The initial vertices \eqref{seq:Gamma0} then simplify to
\begin{subequations}
\label{seq:Gamma0simple}
\begin{align}
\Pi_0(\omega)=&0
\\
S \Gamma_{0}^{+-z}(K_1,K_2,K_3)=&
1,
\\
S^2 \Gamma_{0}^{++--}(K_1,K_2;K_3,K_4)=&
G_0^{-1}(i\omega_3)+G_0^{-1}(i\omega_4)
.
\end{align}
\end{subequations}
Based on these initial conditions, we make the following ansatz for the mixed three-point vertex,
\begin{align}
S\Gamma_\Lambda^{+-z}(K_1,K_2,K_3)
&=
\gamma_\Lambda^0(K_3) 
+
\gamma_\Lambda^1(K_3) G_0^{-1}(\omega_2),
\end{align}
and for the transverse four-point vertex,
\begin{align}
S^2\Gamma_\Lambda^{++--}(K_1,K_2;K_3,K_4)
=&
\nonumber
\\
&\hspace{-2.5cm}
U_\Lambda^{00}(K_1-K_4)
\nonumber
\\
&\hspace{-2.9cm}
+U_\Lambda^{01}(K_1-K_4)G_0^{-1}(\omega_4)
\nonumber
\\
&\hspace{-2.9cm}
+U_\Lambda^{10}(K_1-K_4)G_0^{-1}(\omega_3)
\nonumber
\\
&\hspace{-2.9cm}
+U_\Lambda^{11}(K_1-K_4)G_0^{-1}(\omega_3)G_0^{-1}(\omega_4).
\end{align}
This parametrization retains the initial frequency dependence of the vertices but promotes the corresponding coefficients to $\Lambda$-dependent functions. 
Furthermore, the additional terms $\gamma_\Lambda^1(K)$, $U_\Lambda^{00}(K)$ and $U_\Lambda^{11}(K)$ must necessarily be included since they are generated by the respective flow equation.
With this ansatz, the flow equations~\eqref{seq:gammaflow} may be rewritten in compact matrix form as
\begin{subequations}
\label{seq::flow}
\begin{align}
\partial_\Lambda \Pi_\Lambda(Q)=&\bm{\gamma}_\Lambda^T(Q)\dot{\bm{P}}_\Lambda(Q) \bm{\gamma}_\Lambda (-Q),
\label{seq::piflow}
\\
\partial_\Lambda \bm{\gamma}_\Lambda^T(Q)=&-\bm{\gamma}_\Lambda^T(Q)\dot{\bm{P}}_\Lambda(Q) \bm{U}_\Lambda (Q),
\label{seq::gammaflow}
\\
\partial_\Lambda \bm{U}_\Lambda(Q)=&-\bm{U}_\Lambda(Q) \dot{\bm{P}}_\Lambda(Q) \bm{U}_\Lambda (Q),
\label{seq::uflow}
\end{align}
\end{subequations}
where we introduced the two-component vector
\begin{align}
&\bm{\gamma}_\Lambda(Q)=
\begin{pmatrix}
 \gamma_\Lambda^0(Q)\\
 \gamma_\Lambda^1(Q)
\end{pmatrix},
\end{align}
and the $2\times 2$ matrices
\begin{subequations}
\begin{align}
&\bm{U}_\Lambda(Q)=
\begin{pmatrix}
U_\Lambda^{00}(Q)& U_\Lambda^{01}(Q)
\\
U_\Lambda^{10}(Q)& U_\Lambda^{11}(Q)
\end{pmatrix},
\\
&\dot{\bm{P}}_\Lambda(Q)
=
\begin{pmatrix}
\dot{P}_\Lambda^{00}(Q)& \dot{P}_\Lambda^{01}(Q)
\\
\dot{P}_\Lambda^{10}(Q)& \dot{P}_\Lambda^{11}(Q)
\end{pmatrix}.
\end{align}
\end{subequations}
The four types of generalized differential polarization functions $\dot{P}^{\mu\nu}(Q)$ are defined by
\begin{subequations}
\begin{align}
S^2\dot{P}_\Lambda^{00}(Q) =& 
\int_K 
\dot{L}(K,Q)
\\
S^2\dot{P}_\Lambda^{01}(Q) =& 
\int_K 
\dot{L}(K,Q)
G_0^{-1}(i\omega),
\\
S^2\dot{P}_\Lambda^{10}(Q) =& 
\int_K 
\dot{L}(K,Q)
G_0^{-1}(i\omega-i\nu),
\\
S^2\dot{P}_\Lambda^{11}(Q) =& 
\int_K 
\dot{L}(K,Q)
G_0^{-1}(i\omega)G_0^{-1}(i\omega-i\nu),
\end{align}
\end{subequations}
with the function 
\begin{align}
\dot{L}(K,Q)=\left[ G_\Lambda(K) G_\Lambda(K-Q)\right]^{\bullet}.
\end{align}
The quadratic structure of Eqs.~\eqref{seq::flow} allows us 
to construct a formal solution for the seven functions $\Pi_\Lambda(Q)$, $\gamma_\Lambda^\mu(Q)$ and $U_\Lambda^{\mu\nu}(Q)$.
In particular, 
we note that the flow equation \eqref{seq::uflow} involving the four functions $U_\Lambda^{\mu\nu}(Q)$ is a matrix Ricatti equation, 
whose formal solution is 
\begin{align}
\bm{U}_\Lambda(Q)=\left[ \mathsf{1}+ \bm{U}_0(Q) \bm{P}_\Lambda(Q)\right]^{-1} \bm{U}_0(Q).
\label{seq::usol}
\end{align}
Here, 
$\bm{U}_0(Q)=
\begin{pmatrix}
0 & 1\\1&0
\end{pmatrix}$ is determined via the initial conditions \eqref{seq:Gamma0simple} and the coefficients of the matrix $\bm{P}_\Lambda(Q)$ are given by
\begin{align}
\begin{pmatrix}
\bm{P}_\Lambda(Q)
\end{pmatrix}^{\mu\nu}=P_\Lambda^{\mu\nu}(Q)= \int_{0}^{\Lambda} \mathop{d\Lambda'} \dot{P}^{\mu\nu}_{\Lambda'}(Q).
\label{seq:bubblesint}
\end{align}
Moreover, 
the structure of the flow equations \eqref{seq::flow} allows us to derive an explicit solution for $\bm{\gamma}_\Lambda(Q)$.
Comparing Eqs.~\eqref{seq::gammaflow} and \eqref{seq::uflow}, we can identify the two independent solutions
\begin{equation}
\bm{\gamma}_{\Lambda,1}(Q)=
\begin{pmatrix}
U_\Lambda^{00}(Q)\\U_\Lambda^{01}(Q)
\end{pmatrix},
\end{equation}
and
\begin{equation}
\bm{\gamma}_{\Lambda,2}(Q)=
\begin{pmatrix}
U_\Lambda^{10}(Q)\\U_\Lambda^{11}(Q)
\end{pmatrix}.
\label{seq:gammasol}
\end{equation}
However, only the latter is compatible with the initial condition
$\bm{\gamma}_{\Lambda=0}^T(Q)=\left( 1,0 \right)$ and is thus chosen in the following.
Substituting the solution $\bm{\gamma}_{\Lambda,2}(Q)$ into Eq.~\eqref{seq::piflow} and using the symmetries 
$ U_\Lambda^{00}(-Q)=U_\Lambda^{00}(Q)$ and $U_\Lambda^{01}(-Q)=U_\Lambda^{10}(Q)$, 
it is then straightforward to show that the flow of the polarization satisfies
\begin{align}
\partial_\Lambda \Pi_{\Lambda}(Q)=&
-\partial_\Lambda U_{\Lambda}^{11}(Q).
\label{seq:piurelation}
\end{align}
Integrating both sides of Eq.~\eqref{seq:piurelation} from $\Lambda=0$ to $\Lambda=1$ 
and explicitly evaluating the $11$-matrix element of the solution \eqref{seq::usol} then yields the polarization 
\begin{align}
&\Pi_{\Lambda=1}(Q)=
-U_{\Lambda=1}^{11}(Q)=
\nonumber
\\
&\hspace{.4cm}
\frac{P^{00}(Q)}
{\left[ 1+P^{10}(Q)\right] \left[1+P^{01}(Q)\right] -P^{00}(Q) P^{11}(Q)},
\end{align}
with $P^{\mu\nu}(Q)=P_{\Lambda=1}^{\mu\nu}(Q)$.
Note that upon integrating Eq.~\eqref{seq:piurelation} the initial terms do not contribute since $U_{0}^{11}(Q)=-\Pi_{0}(Q)=0$ for $\nu\neq0$.
Substituting this result into Eq.~\eqref{eq:Gammazz}, we hence obtain the longitudinal susceptibility given in Eq.~\eqref{eq:chire},
\begin{align}
& \chi^{zz}(Q)=
 \nonumber
 \\
 &\frac{P^{00}(Q)}
{\left[ 1+P^{01}(Q)\right] \left[1+P^{10}(Q)\right] -P^{00}(Q) \left[ P^{11}(Q)+J_{\bm{q}}^z \right]}.
\label{seq:chizzrpa}
\end{align}
The four different types of polarizations $P^{\mu\nu}(Q)$ can be further simplified 
by explicitly carrying out the $\Lambda$-integration in Eq.~\eqref{seq:bubblesint}. 
Since the transverse self energy was neglected, 
this is a trivial operation and we recover the generalized polarizations
defined in Eqs.~\eqref{eq:Polres}.

\subsection*{Approximation of the generalized polarizations }
In the following we collect the steps leading to the simplified susceptibility given in Eq.~\eqref{eq:chire2}. 
We assume an isotropic interaction $J_{\bd{k}}^\bot=J_{\bd{k}}^z = J_{\bd{k}} $ from now on, 
and furthermore focus on the regime
$H \lesssim T \ll 1/(2ma^2 )$, where $a$ is the lattice spacing and the magnon mass $m$ 
is defined via the small-momentum expansion of the magnon dispersion,
\begin{equation}
\epsilon_{\bd{k}}=\frac{\bd{k}^2}{2m}.
\end{equation}
We start by carrying out the Matsubara sums in the generalized polarizations defined in Eqs.~\eqref{eq:Polres}.
After partially shifting the loop momentum $\bd{k} = \bd{k}' +\bd{q}/2$  and renaming $\bd{k}' \to \bd{k}$, 
we obtain
\begin{subequations}
\label{seq:bubblesmatsubara1}
\begin{align}
P^{00}(Q) &=
\frac{1}{N}\sum_{\bd{k}} 
\left[
n_{0}' \delta_{\omega,0}
+C(\bd{k},\bd{q},i\nu)
\right],
\\
P^{01}(Q) &= P^{10}(-Q)
\nonumber
\\
&
=
\frac{1}{N}\sum_{\bd{k}} 
\left[
\frac{n_{\bm{k}}-n_{0}}{S} 
+C(\bd{k},\bd{q},i\nu)J_{\bd{k}+\frac{\bm{q}}{2}}
\right],
\\
P^{11}(Q) &=
\frac{1}{N}\sum_{\bd{k}} 
\left[
\frac{2 n_{\bm{k}} J_{\bd{k}}}{S} 
+C(\bd{k},\bd{q},i\nu)
J_{\bd{k}+\frac{\bm{q}}{2}}
J_{\bd{k}-\frac{\bm{q}}{2}} 
\right].
\end{align}
\end{subequations}
Here, 
we introduced the auxiliary function 
\begin{equation}
C(\bd{k},\bd{q},i\nu)=\frac{n_{\bm{k}+\frac{\bm{q}}{2}}-n_{\bm{k}-\frac{\bm{q}}{2}}}{\epsilon_{\bm{k}-\frac{\bm{q}}{2}}-\epsilon_{\bm{k}+\frac{\bm{q}}{2}}+i\nu},
\end{equation}
where
\begin{align}
n_{\bm{k}}&=\frac{1}{e^{\left( H + \epsilon_{\bd{k}} \right)/T }-1}
\end{align}
is the Bose function. 
The initial terms $n_0'=- n_{0} \left( n_{0}+1 \right)/T$ and $n_{0}=\left[e^{ \left( H + S J_0 \right)/T}-1\right]^{-1}$  are exponentially small in the examined temperature range $H\lesssim T$ and are therefore omitted in the following. 
Analyzing the remaining momentum sums 
we note that at low temperatures the thermal momentum 
$k_\text{th} = \sqrt{2mT} \ll 1/a$ acts as ultraviolet cutoff.
In the limit $|\bd{q}| \ll k_\text{th} $, 
we may thus expand the  auxiliary function $C(\bd{k},\bd{q},i\nu)$ as
\begin{align}
C(\bd{k},\bd{q},i\nu)
\approx
c(\bd{k},\bd{q},i\nu)
=
-n'_{\bd{k}}
\frac{\bd{v}_{\bd{k}}\cdot \bd{q} }{\bd{v}_{\bd{k}}\cdot \bd{q}-i\nu}.
\end{align}
Here, 
$n'_{\bm{k}}=-\beta n_{\bm{k}}(n_{\bm{k}}+1)$ and the magnon velocity is $\bd{v}_{\bd{k}}=\nabla_{\bd{k}} \epsilon_{\bd{k}}$.
Moreover, to leading order in $\bd{q}$ the  $\bd{q}$-dependence arising from the factors $J_{\bd{k}\pm \bd{q}/2}$ in Eqs.~\eqref{seq:bubblesmatsubara1} can be neglected because the longitudinal susceptibility \eqref{seq:chizzrpa} depends only on the sum $P^{01} (Q) + P^{10} (Q)$ and the product $P^{01} (Q) P^{10} (Q)$ which are even functions of $\bd{q}$.
We therefore approximate
\begin{subequations}
\label{seq:bubblesmatsubara2}
\begin{align}
P^{00}(Q)  &= 
\frac{1}{N}\sum_{\bd{k}} 
c(\bd{k},\bd{q},i\nu),
\\
P^{10}(Q)&=P^{01}(Q)  
\nonumber
\\
&=\frac{1}{N}\sum_{\bd{k}}
\left[ \frac{n_{\bd{k}}}{S}+
c(\bd{k},\bd{q},i\nu) J_{\bd{k}}\right],
\\
P^{10}(Q)  &= 
\frac{1}{N}\sum_{\bd{k}}
\left[ \frac{2n_{\bd{k}}J_{\bd{k}}}{S}+
c(\bd{k},\bd{q},i\nu) J_{\bd{k}}^2\right].
\end{align}
\end{subequations}
Furthermore, 
in the given temperature range the integrals are dominated by small momenta,
so that we can neglect the momentum dependence of $J_{\bd{k}}$ in the integrands of \eqref{seq:bubblesmatsubara2}. 
Hence, we obtain  
\begin{subequations}
\label{seq:bubblesfinal}
\begin{align}
P^{10}(Q)=&
P^{01}(Q)=
J_0 P^{00}(Q)+\rho,
\\
P^{11}(Q)=& J_0^2 P^{00}(Q)+2 J_0 \rho,
\end{align}
\end{subequations}
for the generalized polarizations in Eqs.~\eqref{eq:Polres} 
where
\begin{equation}
\rho= \frac{1}{NS}\sum_{\bd{k}} n_{\bd{k}}
\label{seq:rho}
\end{equation}
is the density of thermally excited magnons, already defined in Eq.~\eqref{eq:magdens} of the main text.
In order to simplify the numerical analysis it is furthermore convenient to rewrite the integrands of the functions  $P^{00}(Q)$ and $\rho$ in dimensionless form. 
With the substitution $k= k_{\text{th}} \sqrt{\epsilon}$, we obtain
\begin{align}
P^{00}\left(\bd{q},i \omega \right)=P\left(\frac{i\omega}{v_{\text{th}} q}\right),
\label{seq:P00def}
\end{align}
where $v_{\text{th}}=k_{\text{th}}/m$ is the thermal velocity.
The function $P(z)$ has already been defined in Eq.~(8) of the main text; 
i.e.
\begin{align}
P\left(z\right)&=
\frac{\Omega_D (k_{\text{th}} a)^D }{(2\pi)^D2 T}  \int_0^{\infty} \mathop{d\epsilon}  
g_D\left( \frac{z}{\sqrt{\epsilon}}   \right)
\frac{\epsilon^{\frac{D-2}{2}} e^{ h +\epsilon }}{\left( e^{h+\epsilon}-1\right)^2},
 \label{seq:P00dimless}
 \end{align} 
with the dimensionless magnetic field $h=H/T$ 
and the angular average $g_D(z)$ defined in Eq.~\eqref{eq:angularint} of the main text.
Applying the same substitution as above, 
the magnon density $\rho$ in Eq.~\eqref{seq:rho} can be written as 
\begin{align}
\rho=
\frac{\Omega_D (k_{\text{th}}a)^D}{(2\pi)^D2S} \Gamma(D/2) \text{Li}_{\frac{D}{2}}\left(e^{-h} \right),
 \label{seq:rhodimless}
\end{align}
where the polylogarithm $\text{Li}_{s}(z)$ is defined by
\begin{equation}
\text{Li}_{s}(z) = \frac{1}{\Gamma(s)}\int_0^\infty
\mathop{d\epsilon} \frac{\epsilon^{s-1}}{e^{\epsilon}/z-1},
\end{equation}
with the $\Gamma$-function $\Gamma(s)$.
Substituting the approximations \eqref{seq:bubblesfinal} and \eqref{seq:P00def} into Eq.~\eqref{seq:chizzrpa}, 
we finally obtain the expression for the longitudinal susceptibility given in Eq.~\eqref{eq:chire2} of the main text,
\begin{align}
& \chi^{zz}(\bd{q},i\omega) =
\frac{P\left(\frac{i\omega}{v_{\text{th}} q}\right)}
{(1+\rho)^2+J_0 P\left(\frac{i\omega}{v_{\text{th}} q}\right)}.
\label{seq:susceptibility}
\end{align}
Note that in order to be consistent with the previous low momentum expansion, 
we neglected the higher order $\bd{q}$-dependence of $J_{\bd{q}}^z$ in the denominator of Eq.~\eqref{seq:chizzrpa}.

\section*{APPENDIX C: Evaluation of the longitudinal dynamical structure factor in dimensions $D=1,2,3$}
\setcounter{equation}{0}
\renewcommand{\theequation}{C\arabic{equation}}
The longitudinal dynamical structure factor
\begin{align}
&S^{zz}(\bd{q},\omega) = 
\left[1+\frac{1}{e^{\omega/T}-1}\right]\frac{1}{\pi}
\text{Im}  \chi^{zz}\left( \bd{q},\omega +i 0^+ \right) 
\nonumber
\\
& \hspace{-.25cm}
= \frac{1}{1-e^{-\omega/T}}\frac{1}{\pi}
\text{Im}  \left[ 
\frac{P\left(\frac{\omega+i0^+}{v_{\text{th}} q}\right)}
{(1+\rho)^2+J_0 P\left(\frac{\omega+i0^+}{v_{\text{th}} q}\right)}
\right]
\label{eq_s:szz}
\end{align}
may be obtained by evaluating the function $P(z)$ numerically.
The zero-magnon velocity $x_0= \omega_{\bm{q}} /( v_{\text{th}} q ) $ and the damping $\gamma_{\bd{q}}=y_0 v_{\text{th}} q $ can then be obtained from
\begin{equation}
0=(1+\rho)^2+J_0 \text{Re} P(x_0+i0^+),
\label{seq:x0def}
\end{equation}
and
\begin{equation}
y_0=\frac{\text{Im} P(x_0+i0^+)  }{\text{Re}P'(x_0+i0^+) }.
\label{seq:damping}
\end{equation}
To evaluate the integral \eqref{seq:P00dimless}, we need the explicit expressions of the function $g_D(z)$ in dimensions $D=1,2,3$; in particular the real and imaginary parts of the analytical continuation $g_D(x+i0^+)$ with $x \in {\rm I\!R}$.
The analytic properties of this function for arbitrary $D$ are summarized in Ref.~[\onlinecite{Kopietz97}]. 
We obtain in $D=1,2,3$ for the real part
\begin{subequations}
\begin{align}
\text{Re}  g_1\left( x+i0^+ \right)&= 
\begin{cases}
0,& |x|=1
\\
\frac{1}{1-x^2}, & |x|\neq1
\end{cases},
\\
\text{Re} g_2\left( x+i0^+ \right)  &= 
\begin{cases}
1 , & |x|\leq 1
\\
1-\frac{|x|}{\sqrt{x^2-1}}, & |x|>1
\end{cases},
\\
\text{Re}  g_3\left( x+i0^+ \right)  &= 
1-\frac{x}{2} \ln \left| \frac{1+x}{1-x} \right|.
\end{align}
\end{subequations}
Note, that for $x \gg 1$ we may approximate 
\begin{equation}
\text{Re } g_D\left( x+i0^+ \right)\approx -\frac{1}{Dx^2},
\end{equation}
as used in Eq. (18) of the main text.
The imaginary part is given by
\begin{subequations}
\begin{align}
\text{Im}  g_1\left( x+i0^+ \right)  &= 
\frac{\pi}{2} |x| \left[ \delta(1-x)+\delta(1+x) \right],
\\
\text{Im}  g_2\left( x+i0^+ \right)  &= 
 \frac{x}{\sqrt{1-x^2}} \Theta(1-|x|),
\\
\text{Im}  g_3\left( x+i0^+ \right)  &= 
 \frac{\pi}{2} x \Theta(1-|x|).
\end{align}
\end{subequations}
Substituting this into the definition \eqref{seq:P00dimless} yields for the real part of $P(x+i0^+)$
\begin{subequations}
\label{seq:Pre}
\begin{align}
\text{Re}  P_{D=1}(x+i0^+)
&= 
\nonumber
\\
&
\hspace{-2.8cm}
-
\frac{k_{\text{th}} a}{2\pi T}
\mathcal{P} 
\int_{0}^\infty \mathop{d\epsilon} 
\frac{\sqrt{\epsilon}}{(x^2-\epsilon)} \frac{e^{ h+\epsilon }}{\left( e^{h+\epsilon}-1\right)^2},
\\
\text{Re}   P_{D=2}(x+i0^+) 
&= \frac{(k_{\text{th}} a)^2}{4\pi T} 
\Bigg[\frac{1}{e^{h}-1}
\nonumber
\\
& \hspace{-1cm}
- \int_0^{x^2} \mathop{d\epsilon}  
\frac{x }{\sqrt{x^2-\epsilon}} \frac{e^{  h+\epsilon }}{\left( e^{h+\epsilon}-1\right)^2}   \Bigg],
\\
\text{Re} P_{D=3}(x+i0^+) 
&= 
\frac{(k_{\text{th}} a)^3}{4\pi^2 T} 
\Bigg[
\frac{\sqrt{\pi}}{2} \text{Li}_{\frac{1}{2}}\left( e^{- h} \right)
\nonumber 
\\
&
\hspace{-2.8cm}
-\frac{x}{2} \int_0^{\infty} \mathop{d\epsilon}  
\ln \left| \frac{\sqrt{\epsilon}+x}{\sqrt{\epsilon}-x} \right| \frac{e^{  h+\epsilon }}{\left( e^{ h+\epsilon}-1\right)^2 }
\Bigg],
\end{align}
\end{subequations}
where $\mathcal{P}$ denotes the Cauchy principal value.
The imaginary part reads
\begin{subequations}
\label{seq:Pim}
\begin{align}
\text{Im} 
P_{D=1}(x+i0^+) 
&= 
\frac{k_{\text{th}} a}{2 T}
\frac{xe^{  h+x^2}}{\left[ e^{ h+x^2}-1\right]^2},
\\
\text{Im}  P_{D=2}(x+i0^+) 
&= 
 \frac{(k_{\text{th}} a)^2}{4\sqrt{\pi} T} x  
\text{Li}_{-\frac{1}{2}} \left(e^{-h-x^2}\right),
\\
\text{Im}  P_{D=3}(x+i0^+) 
&= 
\frac{(k_{\text{th}} a)^3}{8 \pi T}  \frac{x}{e^{ h+x^2}-1} .
\end{align}
\end{subequations}
The longitudinal structure factor \eqref{eq_s:szz} and the corresponding zero-magnon velocities \eqref{seq:x0def} and damping rates \eqref{seq:damping} 
may now be obtained by calculating the expressions \eqref{seq:Pre} and \eqref{seq:Pim} for $\text{Re }P_D(x+i0^+)$ and $\text{Im }P_D(x+i0^+)$ numerically.

\subsection*{Nearest neighbor exchange}
As emphasized above and in the main text, 
the zero-sound truncation employed in the derivation of the flow equations can formally only be justified for a long-range exchange interaction.
It is nevertheless useful to extrapolate the result \eqref{seq:susceptibility} to a Heisenberg ferromagnet with nearest-neighbor exchange coupling $J$. In this case, we obtain in dimensions $D$ for the interaction $J_0$ and the mass $m$,
\begin{subequations}
\begin{align}
J_0&=2DJ, 
\\
m   =& 1/(  2 J S a^2),
\end{align} 
\end{subequations}
where $a$ is the lattice spacing. 
The thermal momentum and velocity are then given by
\begin{subequations}
\begin{align}
k_{\rm th} =& a^{-1} \sqrt{\frac{T}{JS}},
\\
v_{\rm th} =& \frac{k_{\rm th}}{m}= 2a \sqrt{TJS}.
\end{align} 
\end{subequations}
Note that $J_0=2DJ=D/(Sma^2)$ has also been used to generate Fig.~\ref{fig:dynstruc}.


\begin{thebibliography}{99}
\bibitem{note1}
Note that our expression (\ref{eq:chire}) for the longitudinal susceptibility resembles 
Eq.~(5.13) of Ref.~[\onlinecite{Izyumov02}], 
although it is not equivalent 
because the functions $P^{01}(Q)$, $P^{10}(Q) $ and $P^{11} (Q)$ 
do not agree  with the corresponding quantities introduced in Ref.~[\onlinecite{Izyumov02}].
\end{thebibliography}
\end{document}